\documentclass[10pt,final,journal,twocolumn]{IEEEtran}

\ifCLASSINFOpdf

\else

\fi

\usepackage{cite}
\usepackage[cmex10]{amsmath}
\usepackage{amsmath}
\usepackage{algorithmic}
\usepackage{stfloats}
\usepackage{multicol,multienum}
\usepackage{multirow}
\usepackage[dvips]{graphicx}

\usepackage[tight,footnotesize]{subfigure}
\usepackage{wrapfig}
\usepackage{color}

 \usepackage{booktabs}
 \usepackage{multirow}

\usepackage{mathtools}
\usepackage{color}
\usepackage{float}
\usepackage{bm}

\usepackage{pifont}

\usepackage[ruled,linesnumbered]{algorithm2e}

\usepackage{bigfoot}

\begin{document}

%\title{\huge Many Hands Make Light Work: Leveraging the Power of Ensemble Learning for Secure Low Altitude \textcolor[rgb]{0.00,0.00,0.00}{Economy}}
\title{Leveraging the Power of Ensemble Learning for Secure Low Altitude Economy}
%: Architecture, Challenges, and Future Directions
%: Challenges, Solutions, and Opportunities
%Ensemble Learning for Secure Low Altitude Economics

\author{Yaoqi~Yang,
Yong~Chen,
Jiacheng~Wang,
Geng~Sun, 
Dusit~Niyato,~\IEEEmembership{Fellow,~IEEE},
and Zhu~Han,~\IEEEmembership{Fellow,~IEEE}

\thanks{Manuscript received xxx. The work of Dusit Niyato is supported by Singapore Ministry of Education (MOE) Tier 1 (RG87/22 and RG24/24), the NTU Centre for Computational Technologies in Finance (NTU-CCTF), and the RIE2025 Industry Alignment Fund - Industry Collaboration Projects (IAF-ICP) (Award I2301E0026), administered by A*STAR. (Corresponding author: Geng Sun and Jiacheng Wang)}
\vspace{-0.9cm}

\thanks{Yaoqi~Yang is with the National Key Laboratory on Near-Surface Detection, Beijing, 100072, China. (e-mail: yaoqi$\_$yang@yeah.net)}%Also with the Chengdu Fluid Dynamics Innovation Center, Chengdu 610031, China. 

\thanks{Yong Chen is with the Chengdu Fluid Dynamics Innovation Center, Chengdu 610031, China. (e-mail: literature$\_$chen@nudt.edu.cn)}

\thanks{Jiacheng Wang and Dusit Niyato are with the College of Computing and Data Science, NTU, Singapore. (email: jiacheng.wang@ntu.edu.sg; dniyato@ntu.edu.sg)}

\thanks{Geng Sun is with the College of Computer Science and Technology, Jilin University, Changchun 130012, China, and also with the College of Computing and Data Science, Nanyang Technological University, Singapore 639798 (email: sungeng@jlu.edu.cn).}

\thanks{Zhu Han is with the Department of Electrical and Computer Engineering at the University of Houston, Houston, TX 77004, USA. (e-mail: hanzhu22@gmail.com)}
}

\maketitle

\begin{abstract}
Low Altitude Economy (LAE) holds immense promise for enhancing societal well-being and driving economic growth. However, this burgeoning field is vulnerable to security threats, particularly malicious aircraft intrusion attacks. 
\textcolor[rgb]{0.00,0.00,0.00}{To address the above concerns, intrusion detection systems (IDS) can be used to defend against malicious aircraft intrusions in LAE. Whereas,} due to the heterogeneous data, dynamic environment, and resource-constrained devices within LAE, current IDS face challenges in detection accuracy, adaptability, and resource utilization ratio. \textcolor[rgb]{0.00,0.00,0.00}{In this regard, due to the inherent ability to combine the strengths of multiple models, ensemble learning can realize more robust and diverse anomaly detection further enhance IDS accuracy, \textcolor[rgb]{0.00,0.00,0.00}{thereby} improving robustness and efficiency of the secure LAE. Unlike single-model approaches, ensemble learning can leverage the collective knowledge of its constituent models to effectively defend the malicious aircraft intrusion attacks.} \textcolor[rgb]{0.00,0.00,0.00}{Specifically}, this paper investigates ensemble learning for secure LAE, covering research focuses, solutions, and a case study. We first establish the rationale for ensemble \textcolor[rgb]{0.00,0.00,0.00}{learning and then} review research areas and potential solutions, demonstrating the necessities and benefits of applying ensemble learning to secure LAE. Subsequently, we propose a framework of ensemble learning-enabled malicious aircrafts tracking \textcolor[rgb]{0.00,0.00,0.00}{in the secure LAE}, where its feasibility and effectiveness are evaluated by the designed case study. Finally, we conclude by outlining promising future research directions for further advancing the ensemble learning-enabled secure LAE.
\end{abstract}

%\begin{IEEEkeywords}
%Ensemble learning, Low altitude \textcolor[rgb]{0.00,0.00,0.00}{economy}, Malicious aircraft detection, Object tracking
%\end{IEEEkeywords}

\IEEEpeerreviewmaketitle
\vspace{-0.7cm}

\section{Introduction}

%With the rapid development of social economics, ground space resources are facing serious challenges due to increased building density, vehicle traffic, and overexploited land. To further promote social economics and maintain the sustainability of natural resources, low-altitude economics (LAE) has been proposed. Specifically, LAE refers to the economic activities and opportunities emerging in the airspace within 3000 meters of the ground, primarily driven by the commercial and industrial use of drones and electric vertical take-off and landing (eVTOL) aircraft. However, since LAE operates in an open wireless environment, it remains vulnerable to security concerns stemming from malicious aircraft intrusion attacks [habler2023assessing]. For example, unauthorized drone swarms could disrupt air traffic or deliver dangerous payloads, while compromised eVTOLs could be hijacked and used for malicious purposes. To mitigate these threats and fully leverage the potential of LAE, robust intrusion detection systems (IDS) are crucial. Specifically, an effective IDS can provide real-time monitoring of the low-altitude airspace, detect anomalous flight patterns or unauthorized access attempts, and promptly alert relevant authorities to potential threats, thereby ensuring the safe and secure operation of LAE.

With the rapid development of social economics, ground space resources are facing serious challenges due to the increased buildings, vehicles, and overexploited land. To further promote social \textcolor[rgb]{0.00,0.00,0.00}{economy }and maintain the natural resources sustainability, low altitude economy (LAE) has been proposed. Specifically, LAE refers to the economic activities and opportunities that are emerging in the airspace within 3000m to the ground, mainly driven by the commercial and industrial use of drones and electric vertical take-off and landing (eVTOL) aircrafts \cite{zhou2025cooperative}. However, since LAE \textcolor[rgb]{0.00,0.00,0.00}{covers an} opening wireless environment, it still has security concerns cased by malicious aircrafts intrusion attacks. %\cite{habler2023assessing}
For example, unauthorized drone swarms \textcolor[rgb]{0.00,0.00,0.00}{can} disrupt air traffic near airports, causing significant delays and posing a severe risk of collisions\footnote{https://www.juvare.com/airspace-security-in-2025-addressing-the-rising-risks-of-unauthorized-drones/}. They \textcolor[rgb]{0.00,0.00,0.00}{can} also deliver dangerous payloads, such as explosives or biological agents, to sensitive locations \textcolor[rgb]{0.00,0.00,0.00}{such as} government buildings or critical infrastructure. While compromised eVTOLs \textcolor[rgb]{0.00,0.00,0.00}{can} be hijacked and used for malicious purposes, turning them into airborne weapons or surveillance platforms, posing a direct threat to public safety and privacy. Furthermore, malicious aircrafts \textcolor[rgb]{0.00,0.00,0.00}{can} jam or spoof \textcolor{black}{global positioning system (GPS)} signals of legitimate aircrafts, causing LAE vehicles to deviate from their intended routes, potentially leading to accidents or unauthorized landings in restricted areas.

To mitigate these threats and fully leverage the potential of LAE, robust intrusion detection systems (IDS) can be crucial. Specifically,
\textcolor[rgb]{0.00,0.00,0.00}{intrusion indicates an attacker gains unauthorized access to a device, network, or system with physical or communication/cyber channels. IDS can then observe network traffic for malicious activity and transactions and sends immediate alerts when above-mentioned abnormal actions are observed}\footnote{https://www.geeksforgeeks.org/intrusion-detection-system-ids/}.
An effective IDS can provide real-time monitoring of the low-altitude airspace, detect anomalous flight patterns or unauthorized access attempts, and promptly alert relevant authorities to potential threats, thereby ensuring the safe and secure operation of LAE. By treating the abnormal communication signals or traffic data as traits, current IDS adopts machine learning approach to train \textcolor[rgb]{0.00,0.00,0.00}{an} abnormal LAE signal/data classifier, aiming at identifying malicious aircrafts intrusion attacks with these traits. However, due to the unique \textcolor[rgb]{0.00,0.00,0.00}{characteristics} of LAE, current IDS may suffer from \textcolor[rgb]{0.00,0.00,0.00}{several} challenges:
\begin{itemize}
\item \textcolor{black}{Sparse and heterogeneous data leads to low detection accuracy.} LAE involves diverse types of data, including flight telemetry, sensing data, and environmental information, which \textcolor[rgb]{0.00,0.00,0.00}{could} be sparse and exhibit complex correlations. Traditional IDS often struggle with such heterogeneous and high-dimensional data. 
\item \textcolor{black}{Dynamic and unpredictable environment leads to low adaptability.} The low-altitude airspace is a dynamic environment with rapidly changing weather conditions, traffic patterns, and operational scenarios, making it difficult to establish stable baseline behaviors for intrusion detection.
\item \textcolor{black}{Resource-constrained devices leads to low resource utilization ratio.} Many LAE devices, such as drones, tethered balloon in the sky, have limited \textcolor[rgb]{0.00,0.00,0.00}{communications, computing,} and energy resources, which makes it challenging to implement IDS algorithms.
\end{itemize}

In this regard, ensemble learning, \textcolor[rgb]{0.00,0.00,0.00}{a technique that integrates several base model to improve final outputs, can} enhance the accuracy, robustness, and efficiency of intrusion detection systems in LAE. Therefore, \textcolor[rgb]{0.00,0.00,0.00}{with the} learning's superior generalization capacities and ability to handle complex data, it can offer \textcolor[rgb]{0.00,0.00,0.00}{several benefits to secure LAE}:

\begin{itemize}
\item \textcolor{black}{Improving detection accuracy.} Ensemble learning can combine \textcolor[rgb]{0.00,0.00,0.00}{prediction and estimation results} of multiple models trained on different aspects of the data, thus improving overall detection accuracy and reducing false alarms, even when dealing with noisy and sparse data.
\item \textcolor{black}{Enhancing robustness to dynamic environments.} By leveraging a variety of base models with careful considerations, ensemble learning can adapt to the dynamic and unpredictable nature of the LAE environment and provide more reliable intrusion detection when conditions \textcolor[rgb]{0.00,0.00,0.00}{change rapidly}.
\item \textcolor{black}{Increasing resource utilization efficiency.} Some ensemble learning techniques, such as boosting and bagging, can be implemented in distributed manners with relatively low computational overhead, making them suitable for resource-constrained LAE devices.
\end{itemize}

In this paper, we investigate the motivations, benefits, and potential applications of ensemble learning-enabled secure LAE. Moreover, by answering questions about why, what benefits and how ensemble learning \textcolor[rgb]{0.00,0.00,0.00}{can} help defend against malicious aircrafts intrusion attacks, we evaluate the significance, necessity, importance, and superiority to \textcolor[rgb]{0.00,0.00,0.00}{utilize} ensemble learning \textcolor[rgb]{0.00,0.00,0.00}{for} secure LAE. To be specific, the main contributions can be summarized below.
\begin{itemize}
  \item \textcolor[rgb]{0.00,0.00,0.00}{We introduce} the preliminaries of secure LAE and ensemble \textcolor[rgb]{0.00,0.00,0.00}{learning and explain} the motivations and potential benefits of ensemble learning-enabled secure LAE, which can address concerns faced by current IDS.
  \item \textcolor[rgb]{0.00,0.00,0.00}{We present research focuses of secure LAE and discuss} how to apply ensemble learning to provide possible solutions, mainly including unknown aircrafts detection, identification, localization, and authentication.
  \item To evaluate the performance of ensemble learning-enabled secure LAE, we propose an ensemble learning-based malicious aircrafts tracking framework. Numerical results of the designed case study have demonstrated the \textcolor[rgb]{0.00,0.00,0.00}{accuracy} and effectiveness of the proposal.
\end{itemize}

\textcolor{black}{The rest of the paper is organized as follows. Section II presents preliminaries of secure LAE and ensemble learning. Then, Section III reviews the research focuses and solutions for ensemble learning-enabled secure LAE, where the potential challenges are also discussed. Subsequently, we propose a framework for malicious aircraft tracking with ensemble learning in Section IV, and the designed case study evaluate the proposal's effectiveness. Finally, we outline several open issues in Section V and summarize the paper in Section VI.}
\vspace{-0.5cm}

\section{\textcolor[rgb]{0.00,0.00,0.00}{Overview of Secure Low Altitude Economy and Ensemble Learning}}

%\textcolor[rgb]{0.00,0.00,0.00}{In this section, we first make an overview of secure LAE, presenting its major operations and examples. Then, we introduce the preliminaries of ensemble learning, including its categories and advantages.}

\subsection{\textcolor[rgb]{0.00,0.00,0.00}{Secure Low Altitude Economy}}

\begin{figure*}[!htb]
  \centering
  \includegraphics[width=10.5cm]{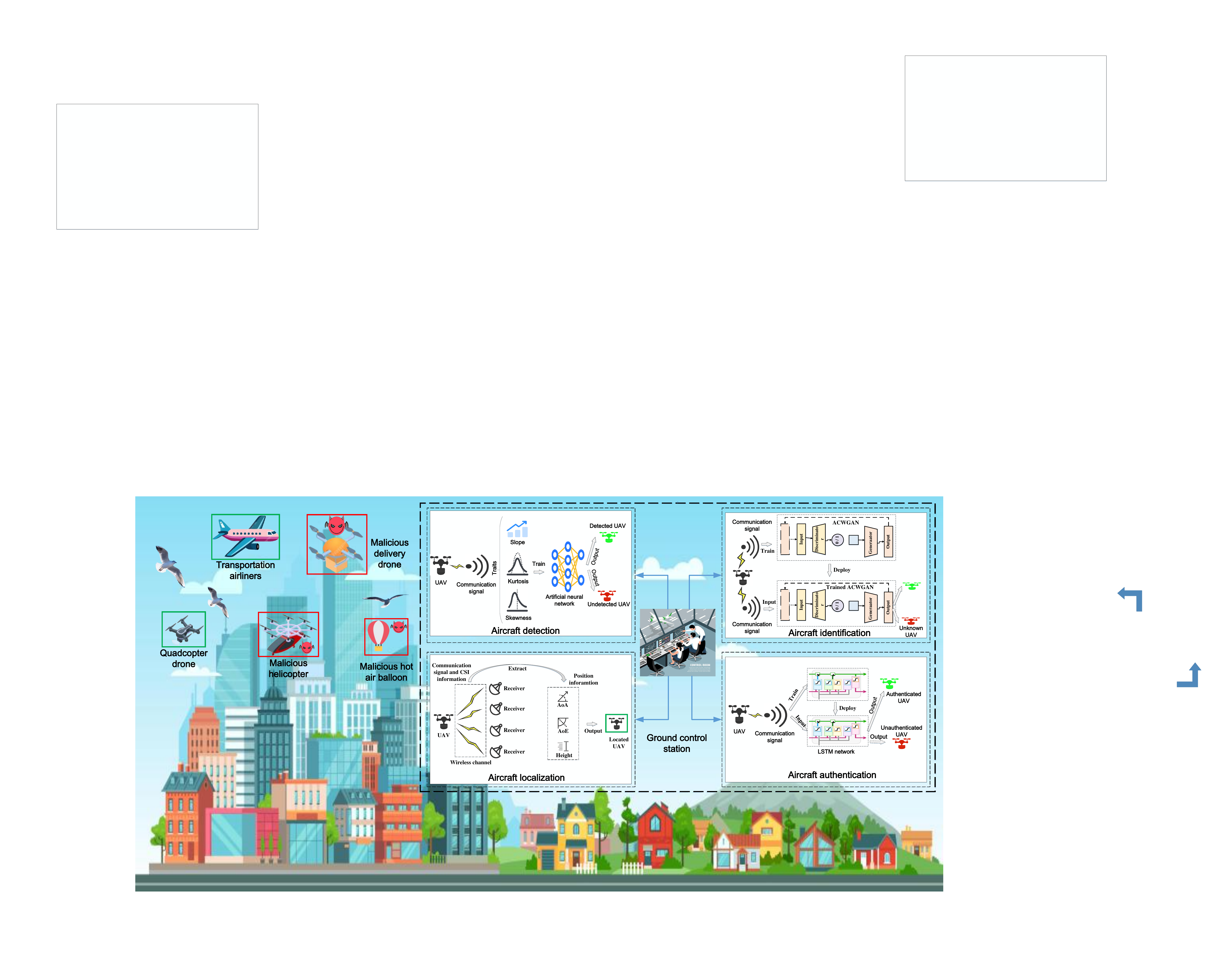}\\
  \caption{\textcolor{black}{The system model of secure LAE, which mainly includes the detection, identification, localization, and authentication of malicious aircrafts.}}\vspace{-0.6cm}
  \label{APPP}
\end{figure*}

As shown in Fig. 1, secure LAE refers to a state where low-altitude airspace operations are protected from unauthorized access, malicious interference, and security breaches, ensuring the safety, reliability, and privacy of all aircrafts and users involved, which facilitates the sustainable growth of the LAE \cite{liu2025research}. To ensure the security of LAE, malicious aircraft intrusion attacks should be carefully defended. In this regard, the major operations that should be conducted for the unknown aircrafts include: aircraft detection, identification, localization, and authentication. The details are presented as follows:
\begin{itemize}
\item Aircraft detection indicates sensing the presence of any aerial vehicle, regardless of its type or intention, within the low-altitude airspace. For example, \cite{zhang2018uav} proposed artificial neural network (ANN)-based detection algorithm for \textcolor[rgb]{0.00,0.00,0.00}{a} \textcolor{black}{unmanned aerial vehicle (UAV)}. By extracting the slope, kurtosis, and skewness features of the communication signal received from the UAV, the proposed \textcolor[rgb]{0.00,0.00,0.00}{algorithm} aims at training the ANN to justify the existence of the target UAV. Experiments results have shown that the proposed \textcolor[rgb]{0.00,0.00,0.00}{algorithm} can reach 82\% detection accuracy, better than active radar and acoustic detection methods, \textcolor[rgb]{0.00,0.00,0.00}{mainly due to its ability to learn complex patterns and discriminate UAV signals from background noise without relying on specific signal signatures}. However, due to the limited feature set and reliance on specific communication signal characteristics, it is difficult to \textcolor[rgb]{0.00,0.00,0.00}{precisely} detect UAVs under \textcolor[rgb]{0.00,0.00,0.00}{high} interference, or employing signal masking techniques.

    %\cite{wu2021improved} proposed an improved Mask R-CNN model to enhance the detection performance in the aircraft remote sensing images. This approach can realize aircrafts detection and segmentation in parallel from the dense targets and complex background, reaching 50.5\% average detection accuracy and 72.3\% mIoU. However, due to the reliance on high-resolution remote sensing image data and the computational complexity of Mask R-CNN, it is difficult to apply this method to real-time and onboard detection systems with limited resources.

\item Aircraft identification represents classifying the detected object as an aircraft, and further determining whether it is a known or unknown entity. For example,
    \cite{zhao2018classification} \textcolor[rgb]{0.00,0.00,0.00}{designed} a UAV classification system using \textcolor[rgb]{0.00,0.00,0.00}{auxiliary classifier Wasserstein generative adversarial networks} (AC-WGANs) based on the wireless signals collected from UAVs of various types. By inputting the data from UAVs, the trained networks can identify the UAV  with its discriminant model for classification. Numerical results have shown that the proposed approach is with a recognition accuracy of around 95\%, outperforming \textcolor[rgb]{0.00,0.00,0.00}{the support vector machine} (SVM) approach. However, due to the reliance on pre-existing knowledge of UAV signal profiles, this approach still faces challenges of accurately identifying \textcolor[rgb]{0.00,0.00,0.00}{unseen} or modified UAV types, and maintaining robustness in dynamic wireless environments.

   % \cite{farhadmanesh2022general} proposed a computer vision-based aircraft identification approach. This approach first identifies the aircraft type with a Convolutional Neural Network (CNN) classifier and then recognizes the tail number in a  probabilistic multi-frame-based (MFB) framework. Numerical results have shown that the proposed approach is with 90\% identification accuracy.     However, due to its dependence on clear, unobstructed visual data and the computational cost of both the CNN and MFB framework, it is difficult to implement in scenarios with poor visibility (e.g., fog, rain) or on devices with limited processing power.

\item Aircraft localization means estimating \textcolor[rgb]{0.00,0.00,0.00}{a} real-time geographic location of a identified aircraft, including its altitude, longitude, and latitude. For example, by monitoring communication signal and channel state information (CSI) of the UAV, \cite{nie2021uav} extracted angle-of-arrival (AOA) and angle of elevation (AOE) information. Then, by combining the \textcolor[rgb]{0.00,0.00,0.00}{line-of-sight} (LoS) of multiple receivers, the UAV's height can be estimated by pitch angle of each receiver, and horizontal and vertical position can be calculated with AOA and AOE values. The experimental results show that  the median accuracy of 2D positioning is 0.76 m, and 3D positioning is 1.2 m. However, the reliance on specific signal features and super-resolution estimation might limit its adaptability to scenarios with significant signal interference or NLoS conditions.

    %based on crowdsourced time of arrival and signal strength measurements data from many different sensors, \cite{adesina2019aircraft} designed a deep neural network (DNN)-based aircraft localization approach, where the aircrafts position can be predicted by DNN as the localization results. Numerical results have shown that the proposal can realize 0.6544, 6.544, and 0.5789 Mean Absolute Percentage Error (MAPE) in latitude, longitude, and altitude aspects. However, due to its reliance on reliable and rich crowdsourced sensor data, it is difficult to ensure consistently accurate localization in areas with sparse sensor coverage.

\item Aircraft authentication refers to confirming aircrafts authorization to operate in the given airspace and adherence to predefined flight plans. For example, \cite{ferdowsi2018deep} proposed a long short term memory (LSTM)-based watermarking  algorithm, enabling the IoT devices (IoTDs) to extract a set of stochastic features from their generated signal and dynamically watermark these features into the signal. By training the LSTM model with above generated features, the proposed approach can effectively authenticate the reliability of the signals. Such an authentication scheme can also be applied to aircrafts authentication by substituting IoTDs' signals into aircraft signals. However, due to the vulnerability to sophisticated signal manipulation attacks that could mimic legitimate watermarks, this method may not be suitable for defending against advanced adversaries with knowledge of the authentication algorithm.

    %\cite{han2023dp} proposed a deep learning-based drone pilot authentication scheme, DP-Authentication, to protect UAVs from malicious radio-manipulated attacks. By inputting UAV flight data into the gated recurrent unit (GRU) classifier-based authentication scheme, the pilot legal status can be validated dynamically. Numerical results have shown that the proposal can authenticate pilots with an accuracy of 95.24\% and detect malicious hijacking with an accuracy of 96.82\%. However, due to the vulnerability of deep learning models to adversarial attacks and the reliance on the integrity of the received UAV flight data, it is difficult to guarantee robust authentication in the face of sophisticated spoofing or manipulation attempts by malicious actors.

\end{itemize}

In summary, current approaches for malicious aircraft detection, identification, localization, and authentication face challenges in model robustness, accuracy, and adaptability. To build a more secure LAE, it is of great significance to adopt ensemble learning to address above concerns.

\vspace{-0.5cm}
\subsection{\textcolor[rgb]{0.00,0.00,0.00}{Ensemble Learning}}

Ensemble learning is a machine learning paradigm that combines the predictions of multiple base models to improve the overall performance and robustness of the system \cite{yang2023survey}. Specifically, ensemble learning leverages the wisdom of the crowd to achieve more robust and accurate predictions, instead of relying on a single, and potentially \textcolor[rgb]{0.00,0.00,0.00}{limited} model. In this regard, ensemble learning is characterized by data processing efficiency enhancement, model accuracy and adaptability improvement. In addition, ensemble methods can be mainly categorized into bagging, boosting, and stacking, the details are listed as follows:
\begin{itemize}

\item \textcolor{black}{The principle of bagging is training multiple base learners on different subsets of the training data, created by random sampling with replacement (bootstrapping).} This approach can reduce variance and over-fitting. For example, the \textcolor[rgb]{0.00,0.00,0.00}{random forest} algorithm uses decision trees as base learners, and then aggregates their predictions through averaging or majority voting to produce a more stable and accurate overall prediction. In this regard, a Random Forest classifier could be trained to identify UAVs based on features extracted from radar signals, acoustic signatures, and video footage. By bagging, the model becomes less susceptible to noise and variations in the data, resulting in more reliable drone detection even in challenging environments \cite{liu2021classification}.

\item \textcolor{black}{Boosting aims at training base learners sequentially, where each subsequent model focuses on the mistakes made by its predecessors, typically by weighting training instances based on their prediction errors.} This approach can reduce bias and improve overall accuracy. The typical applications include AdaBoost, Gradient Boosting Machines (GBM), XGBoost, and CatBoost algorithms \cite{yang2023survey}. Taking an example, boosting algorithms could be used to build a highly accurate UAV identification system based on radio frequency (RF) fingerprinting. By sequentially training models to focus on misclassified UAVs, the system could learn to distinguish between different drone models with high precision \cite{ezuma2019micro}.

\item \textcolor{black}{Stacking focuses on training several base learners, then using their predictions as input to train a meta-learner that combines them.} This approach can improve \textcolor[rgb]{0.00,0.00,0.00}{prediction} accuracy by combining the strengths of different models. For example, we can improve the object classifying accuracy through training multiple diverse models (e.g., a random forest, a support vector machine, and a neural network,) and then use a logistic regression model as the meta-learner to learn how to best weight and combine predictions of these base models. In this regard, for a drone threat assessment system, stacking could be used to integrate information from multiple sensors and analysis techniques, such as radar, acoustic sensors, video analytics, and RF signal analysis, to provide a comprehensive assessment of the threat posed by a drone \cite{alsaffar2024enhancing}.
\end{itemize}

Due to the superiority capacity of ensemble learning, it can be used to address adaptability and robustness concerns of current machine model performances. In this regard, leveraging ensemble learning to help establish a more secure LAE is feasible and with great significance.
\textcolor[rgb]{0.00,0.00,0.00}{Furthermore, there are several learning concepts which are similar to ensemble learning, the details are below. }

\textcolor[rgb]{0.00,0.00,0.00}{1) Fine-tuning\footnote{https://www.ibm.com/think/topics/fine-tuning} leverages a pre-trained model on specific LAE data improves performance, but its reliance on a single, potentially vulnerable model reduces robustness against novel attacks. It risks overfitting to the training data and lacks the diversity for comprehensive threat detection.}

\textcolor[rgb]{0.00,0.00,0.00}{2) Mixture-of-Experts (MoE)\footnote{https://www.ibm.com/think/topics/mixture-of-experts} utilizes specialized subnetworks selected by a gating network, aiming for expertise in particular data regions. However, proper specialization is challenging in the dynamic LAE environment, and the gating network introduces a single point of failure.}

\textcolor[rgb]{0.00,0.00,0.00}{3) Co-training\footnote{https://www.geeksforgeeks.org/what-is-co-training/} leverages diverse ``views'' of data to train multiple models. However, its effectiveness hinges on independent views, which are difficult to achieve in the secure LAE due to highly correlated data streams.}

\textcolor[rgb]{0.00,0.00,0.00}{4) Knowledge distillation\footnote{https://www.ibm.com/think/topics/knowledge-distillation} can distill knowledge from a complex ``teacher'' model to a smaller ``student'' offers efficiency, but inherently sacrifices robustness and diversity, making it less suitable for the high-security demands of LAE.}

\textcolor[rgb]{0.00,0.00,0.00}{5) Transfer learning\footnote{https://www.ibm.com/think/topics/transfer-learning} addresses limited data by applying knowledge from related problems. While valuable for initializing models, it lacks the inherent diversity, robustness, and anomaly detection capabilities essential for securing the LAE.}

%Note that while both ensemble Learning and mixture of experts (MoE) integrate multiple models, ensemble learning stands apart by its deliberate construction of diverse base learners, frequently trained on distinct \textcolor[rgb]{0.00,0.00,0.00}{subsets of data}, with the express goal of enhancing generalization and robustness across the entire input space. MoE, conversely, centers on specialization, utilizing ``expert" sub-networks tailored to particular regions or facets of the input, dynamically selected or weighted by a gating network. 

Based on above analysis, \textcolor{black}{ensemble learning stands out as the most suitable approach for securing the LAE among other learning schemes. To be detailed, the strength of ensemble learning resides in its general-purpose capacity to boost predictive accuracy and stability through the synthesis of diverse insights, positioning it as a vital methodology for applications necessitating consistent and reliable predictions in the face of wide-ranging and intricate uncertainties. Ensemble learning's broader generalization, across many more types of scenarios, often makes it a better choice than above learning schemes.}
Besides, the above three major ensemble learning models are compared in Table I, where their architecture, principle, advantages, disadvantages, and applications in secure LAE are presented.

\begin{table*}[t]
\centering
\caption{\textcolor{black}{Summarization and comparison for the typical ensemble learning models.}} \vspace{-0.2cm}
\label{Summarization}
\resizebox{0.9\width}{!}{
\begin{tabular}{c}
\includegraphics[width=14cm]{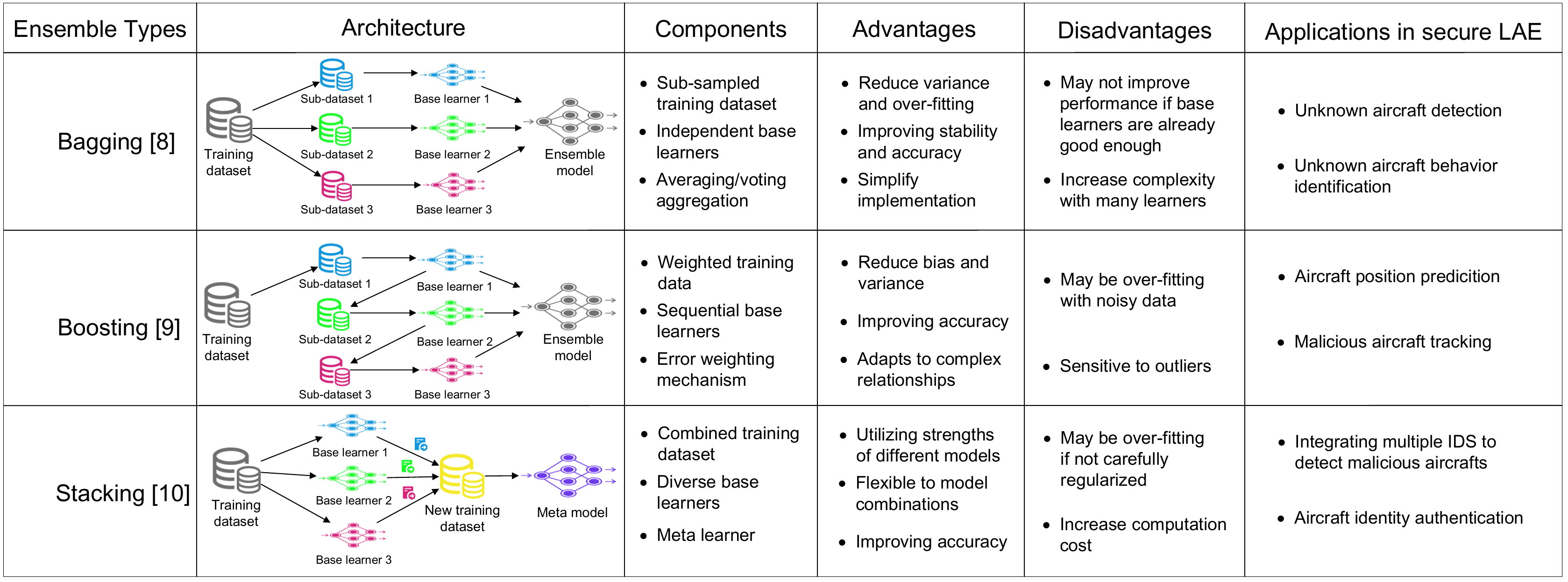}\\
\end{tabular}}
\end{table*}

\vspace{-0.5cm}
\section{Ensemble Learning for Secure LAE}

%\textcolor[rgb]{0.00,0.00,0.00}{In this section, we first introduce the research focuses and solutions of the secure LAE. Then, we summarize the benefits of ensemble learning and potential challenges.}

\subsection{Research Focuses and Solutions}

%\cite{lu2023adversarial}

%\cite{khan2023using}

\subsubsection{Ensemble Learning for Aircrafts Detection}\

Current IDS approaches, such as camera-based detection, and radar-based detection, can be used to detect aircrafts in LAE. However, they face challenges in handling the unique characteristics of the low-altitude environment, including its dynamic nature, and the heterogeneity of data. In this regard, ensemble learning can offer significant insights by leveraging its ability to combine the strengths of multiple diverse models, enhance robustness, and adapt to evolving environments, thus improving the accuracy and reliability of intrusion detection in LAE. For example, \cite{singh2023ensemble} developed an \textcolor[rgb]{0.00,0.00,0.00}{ensemble-based IoT-enabled drones detection scheme} (EDDSBS) to enhance community safety. EDDSBS can be divided into three parts: At first, the aerial image is input to AlexNet model to achieve background substraction. Then, ResNet50 model and YOLOv4 model are combined in the bagging ensemble manner to deal with the processed image by AlexNet model. At last, the output feature vectors from ResNet50 model and YOLOv4 model are simultaneously passed through a detection module to classify the image as a drone or not. \textcolor[rgb]{0.00,0.00,0.00}{By} leveraging the ensemble learning strategy, EDDSBS can integrate ResNet50 and YOLOv4 models with background subtraction to achieve higher drone detection accuracy. The results demonstrated that ensemble learning-based approach can realize a 91.20\% accuracy rate, a significant improvement over a transfer learning-only approach, which achieved 89.32\%.

\subsubsection{Ensemble Learning for Aircrafts Identification}\

IDS schemes including signature-based identification, and behavior-based identification approaches, can be used to identify aircrafts in LAE. However, they face challenges in dealing with the sophisticated camouflage techniques of malicious aircraft, and the variability of flight patterns. \textcolor[rgb]{0.00,0.00,0.00}{Ensemble} learning can offer significant insights by enhancing the robustness of identification through combining diverse models that focus on different features, thereby improving the identification accuracy and reducing false positives even when malicious aircraft employ deceptive tactics. For instance, \cite{alkadi2021identifying} explored an ensemble learning-based identification approach for drones as a defense against hijacking. Specifically, the inertial measurement units (IMU) and radio control (RC) signals can be obtained from the drone's onboard sensors. Then, according to the bagging-based ensemble learning framework, these signals are used for training classify models, where different LSTM classifier are trained with different IMU and RC signals sub-dataset. Finally, the sliding window is utilized to perform majority voting on the output from various classifiers to determine the final identification results. \textcolor[rgb]{0.00,0.00,0.00}{A} bagged ensemble learner with 30 trees outperforms SVM, k-nearest neighbor (KNN), and linear discriminant analysis, reaching a dronep identification accuracy of 90.1\%.

%This study involved analyzing fifteen inertial measurement unit (IMU) readings and four radio control signals to identify the pilot. Their research indicated that a sequence classification scheme utilizing frequency-domain features and an ensemble of random trees could achieve

\subsubsection{Ensemble Learning for Aircrafts Localization}\

Recent developed approaches, like single-sensor localization, multi-sensor fusion localization, can be used to localize aircrafts in LAE. However, they face challenges in achieving high precision in dynamic environments, mitigating signal interference and multi-path effects. In this regard, ensemble learning can offer significant insights by integrating information from multiple localization methods, improving localization accuracy and robustness through model averaging and error correction. For example, in a multi-sensor fusion system, \cite{dasika2020ensemble} developed an ensemble learning approach to improve the accuracy of tracking flights across India. The proposed ensemble learning-based scheme includes three major steps. Firstly, based on the multi-sensor fusion data, an ensemble model of logistic regression and XGBoost are trained to predict the splitting error. Then, XGBoost is used to predict the merging error. Finally, the nearest neighbour search is adopted to compensate for the predicted errors by retaining the data points removed in the first two steps while maintaining the tracking accuracy. In this regard, the proposed ensemble learning-based approach can predict splitting and merging of tracks, addressing merging and splitting errors caused by asynchronous radar observations, along with nearest neighbor search to compensate for identified errors. Results have shown that the tracking accuracy can be improved from 78\% to 93\%.

\subsubsection{Ensemble Learning for Aircrafts Authentication}\

Cryptographic authentication and access control mechanisms can \textcolor[rgb]{0.00,0.00,0.00}{authenticate and prevent} malicious aircrafts in LAE. However, they face challenges in addressing the complexities of key management in a dynamic LAE environment, and the limitations of centralized access control systems in handling a large number of heterogeneous devices and operators. In this regard, ensemble learning can offer significant insights by combining multiple authentication methods to improve resilience against attacks, adapting to varying operational conditions through dynamic adjustments, and enhancing overall authentication robustness via multi-layered validation from diverse perspectives. For example,
RF fingerprinting (RFF) is a non-encrypted  authentication technique that provides an additional layer of security for drones,
\cite{zheng2024convolutional} investigated the CNN and ensemble learning-based UAV authentication scheme with RFF data. This approach first collects the UAV communication signals to establish the RFF dataset for training the authentication model. Then, different sub-datasets are used to train the CNN-based RFF identification models. Finally, various identification results from different CNN models are averaged to determine the final authentication results. Numerical results have shown that the proposed ensemble learning-based method can reached a authentication accuracy of 97.43\%.

%\cite{salunke2023ensemble} explored ensemble learning to improve continuous user authentication in real-world environments. Based on the collected keystroke data from 48 users, the ensemble learning with a soft-voting manner can increase the cumulative accuracy to 24.03\%, outperforming 1D CNN and TabNet models.

\subsection{Lessons learned}

%\begin{table*}[!htb]
%\centering
%\caption{Comparison of traditional approaches and ensemble learning methods in the secure LAE.}
%\resizebox{1.0\width}{1.0\height}{
%\begin{tabular}{c}\label{SYSTEMMODEL33}
%\includegraphics[width=17cm]{comparison4application.pdf}\\
%\end{tabular}}
%\end{table*}

By defending against malicious aircraft intrusion attacks through unknown aircrafts detection, identification, localization, and authentication aspects, ensemble learning can help establish a secure LAE. %\textcolor[rgb]{0.00,0.00,0.00}{Its primary strength lies in its robustness and resilience. By combining multiple diverse models, it mitigates the weaknesses of any single model, making it significantly harder for adversaries to evade detection. This diversity also allows ensembles to effectively handle the complex and dynamic data streams inherent in the LAE, enabling superior anomaly detection and real-time threat assessment. The inherent redundancy within an ensemble provides fault tolerance, if one model fails or is compromised, the system can continue to operate effectively. Therefore, ensemble learning can offer adaptability, readily incorporating new models to respond to evolving threats.}
%A comprehensive comparison of traditional approaches and ensemble learning methods are summarized in Table II.
However, despite ensemble learning has lots of advantages, integrating ensemble learning into secure LAE also faces several challenges:
\begin{itemize}
\item High computational complexity. Ensemble methods often involve multiple models, potentially increasing computational complexity.

\item Data imbalance and scarcity: Training effective ensemble models requires a large and well-balanced dataset representing both benign and malicious aircraft behaviors.  %However, data on malicious aircraft intrusions in LAE is inherently scarce and often imbalanced (many more benign than malicious events). This data imbalance can lead to biased models and reduced performance in detecting rare, but potentially critical, malicious activities. Furthermore, the high dimensionality and heterogeneous nature of LAE data (sensor readings, flight plans, communication logs, etc.) further complicate the task of effective data acquisition and preprocessing.

\item Model interpretability. Deep learning-based ensemble learning models, can be difficult to understand how they arrive at their decisions.  %This lack of interpretability can hinder trust and acceptance, particularly in safety-critical applications like LAE security. Understanding why a system flagged a particular aircraft as malicious is crucial for building confidence and for debugging. Without explainability, identifying and rectifying errors becomes challenging.
\end{itemize}

\section{Ensemble Learning-based Malicious Aircrafts Tracking in Low Altitude \textcolor[rgb]{0.00,0.00,0.00}{Economy}}

%\textcolor[rgb]{0.00,0.00,0.00}{In this section, we first propose a framework for malicious aircraft tracking with ensemble learning. Then, the designed case study and numerical results evaluate the proposal's correctness and effectiveness.}

\subsection{Research Motivation}
%\cite{yang2021security}
\textcolor{black}{Malicious aircrafts tracking encompasses simultaneous detection, identification and localization for the unauthorized/illegal aircrafts.} To achieve this goal, a computer vision-based image detector can be the alternative, which can be mainly divided into one-stage and two-stage categories. One-stage detectors, such as the YOLOX detector, excel in speed and real-time processing, but face challenges in achieving high accuracy, particularly with small objects, and struggle with precise localization. Conversely, two-stage detectors, \textcolor[rgb]{0.00,0.00,0.00}{such as} Fast R-CNN, offer superior detection accuracy and perform better on small or occluded \textcolor[rgb]{0.00,0.00,0.00}{objects. However, it has concerns on the real-time application and high computational resource demands}. \textcolor{black}{Therefore, our goal is to achieve both high accuracy and real-time performance for malicious aircraft tracking. Integrating these two detector types using an ensemble learning approach leverages their complementary strengths, improving overall accuracy, robustness, and real-time capabilities while mitigating their potential individual weaknesses.}
\vspace{-0.5cm}
\subsection{Proposed Framework}

\begin{figure*}[!htb]
  \centering
  \includegraphics[width=10cm]{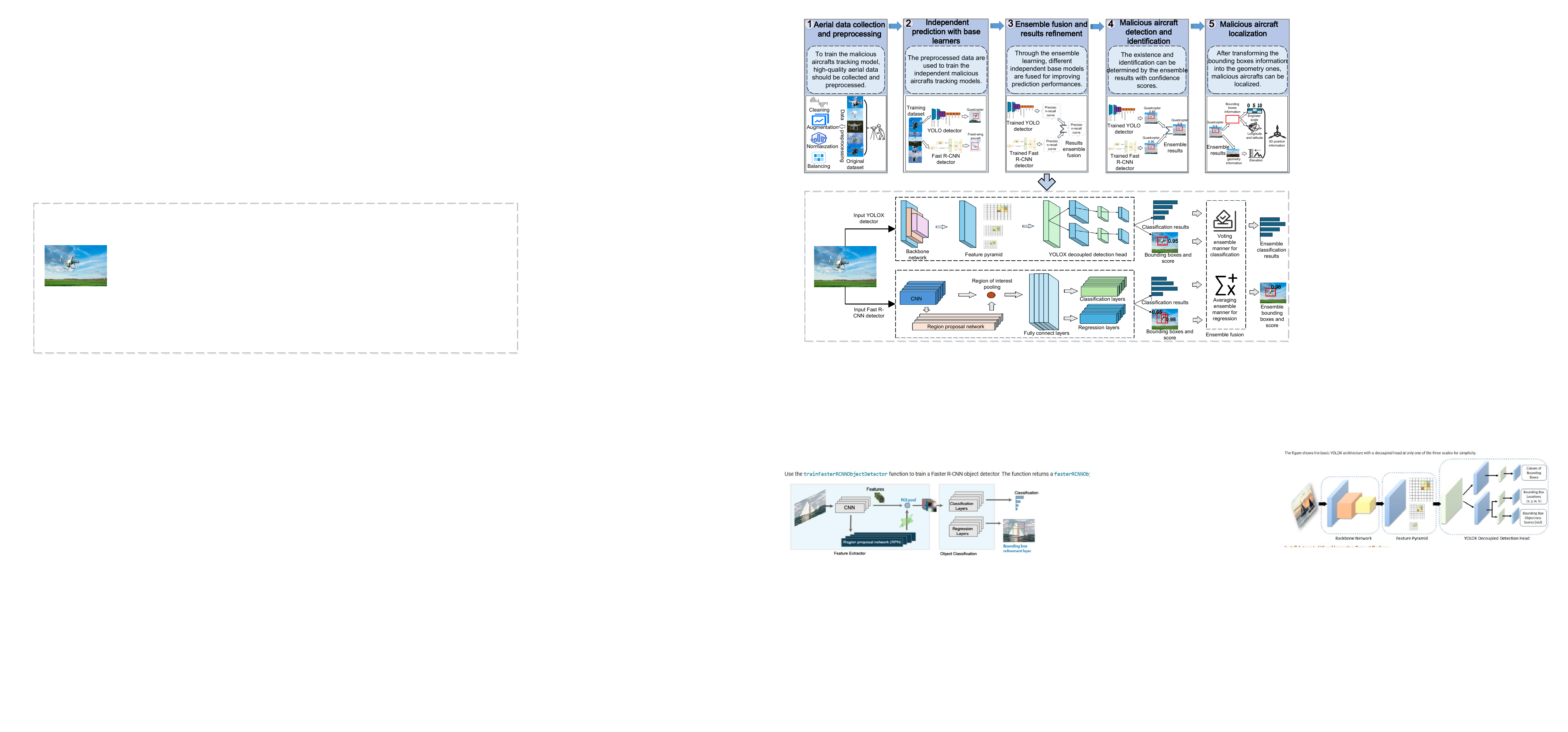}\\
  \caption{The framework for the malicious aircraft tracking with ensemble learning in secure LAE. The ensemble model consists of the YOLOX and Fast R-CNN detectors. It can not only detect and identify the malicious aircrafts with confidence scores, but also localize the malicious aircrafts with the predicted bounding box information.}\vspace{-0.7cm}
  \label{APPP}
\end{figure*}

Based on the ensemble learning approach, malicious aircraft can be detected, identified, and localized in a timely and accurate manner. Specifically, as shown in Fig. 2, we propose a framework for the malicious aircraft tracking with ensemble learning in secure LAE, which mainly includes the following five major steps.

\textbf{Step 1: Aerial Data Collection and Preprocessing.} To provide the input data for detector training and further implementation, the real-time and high quality aerial data \textcolor[rgb]{0.00,0.00,0.00}{is supplied}. \textcolor[rgb]{0.00,0.00,0.00}{Moreover}, due to the scarcity of malicious aircraft samples, the dataset may face imbalance concerns. In this regard, preprocessing techniques \textcolor[rgb]{0.00,0.00,0.00}{such as} noise reduction, image enhancement and data augmentation should be applied to improve image quality and sample category.

\textbf{Step 2: Independent Prediction with Base Learners.} The preprocessed aerial data \textcolor[rgb]{0.00,0.00,0.00}{is} then fed into the detectors, e.g., YOLOX and Fast R-CNN models, serving as base learners. Each model independently predicts potential malicious aircraft class probabilities and locations, generating a set of confidence scores and bounding boxes. Specifically, different base learners are with various advantages. For example, YOLOX excels at quickly identifying potential regions of interest across the entire image, while Fast R-CNN provides more precise bounding box regression and classification within those regions.

\textbf{Step 3: Ensemble Fusion and Results Refinement.} To capitalize on the complementary strengths of base learners, e.g., YOLOX and Fast R-CNN, their predictions are fused using a weighted averaging technique. This approach dynamically adjusts the influence of each model based on the confidence of their individual detections. For example, Fast R-CNN's high-confidence detections are given greater weight due to its superior accuracy, while YOLOX's predictions are emphasized in cases where both models exhibit low confidence or disagreement, leveraging its faster processing to ensure timely results. Subsequently, non-maximum suppression is applied to eliminate redundant bounding boxes and refine the output.

\textbf{Step 4: Malicious Aircraft Detection and Identification.} The refined bounding boxes are analyzed to determine if they contain malicious aircraft as well as their types. To be specific, based on the ensemble results, for the aircrafts marked by the bounding boxes, they are regarded as the malicious aircrafts, and vice versa. Additionally, their categories can simultaneously identified with the ensemble confidence scores.

\textbf{Step 5: Malicious Aircraft Localization.} \textcolor{black}{The refined bounding box coordinates from the previous steps are transformed into geographical coordinates (latitude, longitude, altitude) to accurately pinpoint the location of the detected malicious aircraft in the LAE.} This transformation utilizes camera parameters and altitude data to project the bounding box onto the Earth's surface.

\vspace{-0.5cm}
\subsection{Case Study and Numerical Results}

\begin{figure}[!htb]
\centering
\subfigure[]
{
    \label{systemmodel2}
    %newdatamax2
    \includegraphics[width=3.95cm]{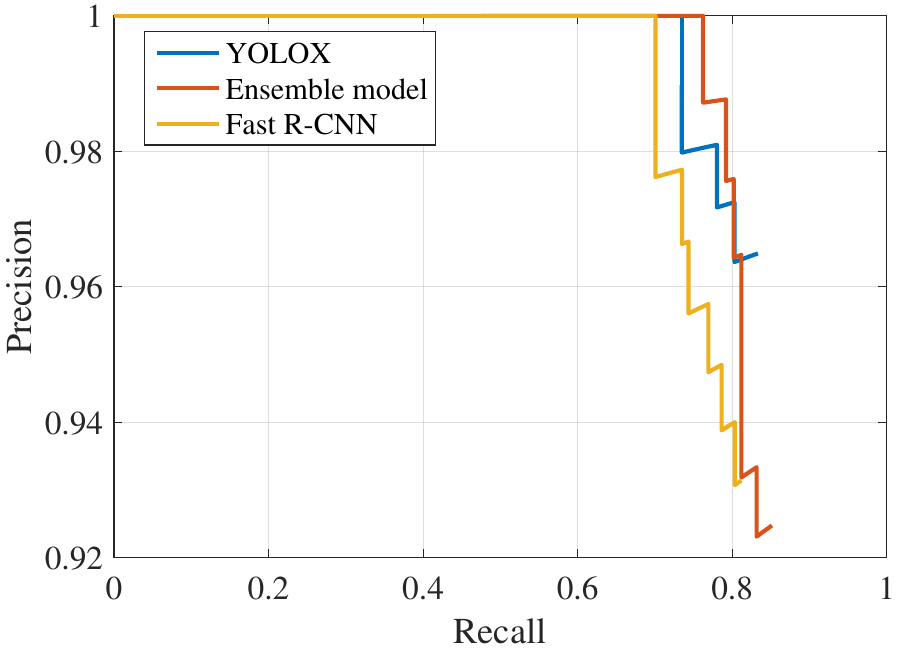}
}
\quad
\hspace{-0.3in}
\subfigure[]
{
   \label{systemmodel3}
   %newdataave1
    \includegraphics[width=3.95cm]{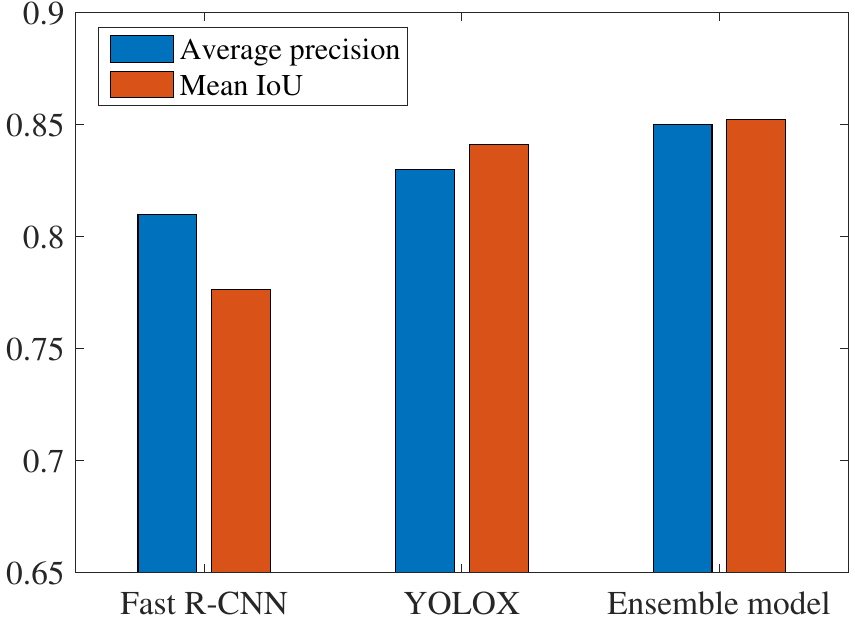}
}
\quad
\hspace{-0.3in}
\caption{Performance evaluation of ensemble model and individual detectors. (a) Precision-recall curve of the individual and ensemble detectors. (b) Average precision and mean IoU performances of the individual and ensemble detectors.}\vspace{-0.3cm}
\label{systemmodel23}
\end{figure}

%\cite{mccoy2023ensemble}

\textbf{Scenario and goal}. In this case study, based on the remote sensing aerial images from ground observation station, we \textcolor[rgb]{0.00,0.00,0.00}{focus on }the malicious aircrafts tracking by ensemble fusion of YOLOX and fast R-CNN detectors. Specifically, we assume that the aircrafts have been recognized as the illegal ones by other investigative techniques (e.g., flight path analysis, communication interception, or pre-flight checks) in advance. The goal is to accurately detect the existence of these flagged malicious aircraft, identify their categories (e.g., drone, eVTOL, or helicopter), and predict their geographic locations (latitude, longitude, and altitude) with high precision. \textcolor[rgb]{0.00,0.00,0.00}{By} comparing with \textcolor[rgb]{0.00,0.00,0.00}{the baseline} individual detector's performances, we aim to demonstrate the effectiveness of the ensemble approach in providing timely and accurate situational awareness within the LAE.

%\textbf{Parameter settings}. The simulation experiments are based on the YOLOX\footnote{https://www.mathworks.com/help/vision/ug/getting-started-with-yolox-object-detection.html} and Fast R-CNN\footnote{https://www.mathworks.com/help/vision/ug/getting-started-with-r-cnn-fast-r-cnn-and-faster-r-cnn.html} models, which are implemented on Matlab R2024b. The ensemble manner is bagging. In addition, the dataset of aircraft images is from \cite{pawelczyk2020real}, which is available on github\footnote{https://github.com/Maciullo/DroneDetectionDataset}.  The training and evaluation operations for the above models are executed on a laptop with AMD Ryzen 9 5900HX CPU @ 3.30GHz, NVIDIA GeForce RTX 3060 GPU, 32.0 GB memory and Windows 11 home operation system.

\textbf{Implementation processes}. At first, we train the YOLOX and Fast R-CNN detectors \textcolor{black}{with different sub-training dataset \cite{pawelczyk2020real} respectively.} Then, we determine the fusion weights for the ensemble model, which is calculated by the average precision ration of YOLOX and Fast R-CNN detectors. Subsequently, we evaluate the performance of ensemble model on precision-recall curve, average precision and mean IoU values. The precision-recall curve provides a comprehensive view of the model's ability to identify positive instances across different confidence thresholds, balancing the trade-off between \textcolor{black}{precision (the accuracy of positive predictions) and recall (the ability to find all positive instances).} Average precision summarizes the precision-recall curve into a single metric, representing the average precision achieved across all recall levels for a specific class. \textcolor{black}{Mean Intersection over Union (IoU) measures the accuracy of bounding box localization by calculating the average overlap} between predicted and ground truth bounding boxes across all instances and classes.

\textbf{Numerical results}. Specifically, as shown in Fig. 3(a), since the ensemble model leverages the complementary strengths of both YOLOX (known for its speed and ability to detect small objects) and Fast R-CNN (known for its high precision and robustness to complex scenes), it achieves a more favorable balance between precision and recall across a wider range of confidence thresholds. The ensemble model can reach the best precision-recall performance than YOLOX and Fast R-CNN. In addition, in Fig. 3(b), the average precision values of Fast-RCNN, YOLOX, and ensemble model are 0.81, 0.83, and 0.85, respectively. The mean IoU values of Fast-RCNN, YOLOX, and ensemble model are 0.7762, 0.8411, and 0.815, respectively. Above results can all evaluate the effectiveness of the ensemble model, which can help better track the malicious aircrafts in the secure LAE.

\vspace{-0.2cm}
\section{Open Issue}

\vspace{-0.2cm}
\subsection{\textcolor{black}{Real-Time Performance in Malicious Aircraft Tracking}}
\textcolor{black}{While ensemble learning offers benefits for malicious aircraft tracking, a key consideration is achieving real-time responsiveness, a critical requirement in this dynamic environment. In this regard, a promising direction involves exploring methods to enhance real-time processing efficiency, such as model compression, parallelization, and edge resource optimization.}
\vspace{-0.5cm}
\subsection{Integration with Other Alarm techniques}
Ensemble learning can still generate false positives that require \textcolor[rgb]{0.00,0.00,0.00}{further enhancement}. Integrating ensemble learning with other alarm techniques, such as pre-filtering alarms with rule-based systems or incorporating contextual information (e.g., weather conditions, \textcolor[rgb]{0.00,0.00,0.00}{and scheduled} events), can help to reduce the number of false alarms and improve the system efficiency.
\vspace{-0.5cm}
\subsection{Trust in Decision-Making}
\textcolor{black}{It is crucial to develop explainable AI (XAI) techniques that can provide insights into the decision-making process and privacy protection of ensemble models. This includes identifying the key features and base learners that contribute to specific predictions, quantifying uncertainty, and providing justifications for actions taken (e.g., alerting authorities or initiating countermeasures).}
\vspace{-0.5cm}
\section{Conclusion}

In this paper, we \textcolor[rgb]{0.00,0.00,0.00}{have introduced }ensemble learning to secure LAE in malicious aircrafts intrusion defense. This combination can significantly address heterogeneous data, dynamic environment, and resource-constrained devices concerns within LAE. \textcolor[rgb]{0.00,0.00,0.00}{We have first presented} the preliminaries of secure LAE and ensemble learning. Then, \textcolor[rgb]{0.00,0.00,0.00}{we have introduced} how to apply ensemble learning to secure LAE, by reviewing research focuses and solutions from malicious aircrafts detection, identification, localization, and authentication aspects. Subsequently, \textcolor[rgb]{0.00,0.00,0.00}{we have proposed} a framework of ensemble learning-enabled malicious aircrafts tracking, with a case study demonstrating its effectiveness. \textcolor[rgb]{0.00,0.00,0.00}{Finally, several open issues are
discussed. We hope this paper can offer insights to establish the secure LAE.}

\vspace{-0.5cm}
\bibliography{ref}{}

@article{singh2023ensemble,
  title={An Ensemble-Based IoT-Enabled Drones Detection Scheme for a Safe Community},
  author={Singh, Jaskaran and Sharma, Keshav and Wazid, Mohammad and Das, Ashok Kumar and Vasilakos, Athanasios V},
  journal={IEEE Open Journal of the Communications Society},
  volume={4},
  number={4},
  pages={1946--1956},
  year={Aug. 2023},
  publisher={IEEE}
}

@article{alkadi2021identifying,
  title={Identifying drone operator by deep learning and ensemble learning of imu and control data},
  author={Alkadi, Ruba and Al-Ameri, Sultan and Shoufan, Abdulhadi and Damiani, Ernesto},
  journal={IEEE Transactions on Human-Machine Systems},
  volume={51},
  number={5},
  pages={451--462},
  year={Sep. 2021},
  publisher={IEEE}
}

@inproceedings{dasika2020ensemble,
  title={An Ensemble Learning Approach to Improve Tracking Accuracy of Multi Sensor Fusion},
  author={Dasika, Anoop Karnik and Paruchuri, Praveen},
  booktitle={Neural Information Processing: 27th ICONIP},
  pages={704--712},
  year={Nov. 2020},
  address={Bangkok, Thailand},
  organization={}
}

@article{liu2025research,
  title={Research on the Security Risk Governance Roadmap in Low-Altitude Economic Field Based on the Economic Externality Theory},
  author={Liu, Shuang and Liu, Mingming},
  journal={Engineering Proceedings},
  volume={80},
  number={1},
  pages={14--30},
  year={Jan. 2025},
  publisher={MDPI}
}

@article{yang2023survey,
  title={A survey on ensemble learning under the era of deep learning},
  author={Yang, Yongquan and Lv, Haijun and Chen, Ning},
  journal={Artificial Intelligence Review},
  volume={56},
  number={6},
  pages={5545--5589},
  year={Nov. 2022},
  publisher={Springer}
}

@article{pawelczyk2020real,
  title={Real world object detection dataset for quadcopter unmanned aerial vehicle detection},
  author={Pawe{\l}czyk, Maciej and Wojtyra, Marek},
  journal={IEEE Access},
  volume={8},
  pages={174394--174409},
  year={Sep. 2020},
  publisher={IEEE}
}

@article{zhang2018uav,
  title={A UAV detection algorithm based on an artificial neural network},
  author={Zhang, Hao and Cao, Conghui and Xu, Lingwei and Gulliver, T Aaron},
  journal={IEEE Access},
  volume={6},
  pages={24720--24728},
  year={May. 2018},
  publisher={IEEE}
}

@inproceedings{zhao2018classification,
  title={Classification of small UAVs based on auxiliary classifier Wasserstein GANs},
  author={Zhao, Caidan and Chen, Caiyun and Cai, Zhibiao and Shi, Mingxian and Du, Xiaojiang and Guizani, Mohsen},
  booktitle={IEEE Global Communications Conference (GLOBECOM)},
  pages={206--212},
  year={Dec. 2018},
  address={Abu Dhabi, United Arab Emirates},
  organization={}
}

@article{nie2021uav,
  title={UAV detection and localization based on multi-dimensional signal features},
  author={Nie, Wei and Han, Zhi-Chao and Li, Yi and He, Wei and Xie, Liang-Bo and Yang, Xiao-Long and Zhou, Mu},
  journal={IEEE Sensors Journal},
  volume={22},
  number={6},
  pages={5150--5162},
  year={Aug. 2021},
  publisher={IEEE}
}

@article{ferdowsi2018deep,
  title={Deep learning for signal authentication and security in massive internet-of-things systems},
  author={Ferdowsi, Aidin and Saad, Walid},
  journal={IEEE Transactions on Communications},
  volume={67},
  number={2},
  pages={1371--1387},
  year={Oct. 2018},
  publisher={IEEE}
}

@article{liu2021classification,
  title={Classification of bird and drone targets based on motion characteristics and random forest model using surveillance radar data},
  author={Liu, Jia and Xu, Qun Yu and Chen, Wei Shi},
  journal={IEEE Access},
  volume={9},
  pages={160135--160144},
  year={2021},
  publisher={IEEE}
}

@inproceedings{ezuma2019micro,
  title={Micro-UAV detection and classification from RF fingerprints using machine learning techniques},
  author={Ezuma, Martins and Erden, Fatih and Anjinappa, Chethan Kumar and Ozdemir, Ozgur and Guvenc, Ismail},
  booktitle={IEEE Aerospace Conference},
  pages={},
  year={Jun. 2019},
  address={Big Sky, MT},
  organization={}
}

@article{alsaffar2024enhancing,
  title={Enhancing intrusion detection systems with dimensionality reduction and multi-stacking ensemble techniques},
  author={Alsaffar, Ali Mohammed and Nouri-Baygi, Mostafa and Zolbanin, Hamed},
  journal={Algorithms},
  volume={17},
  number={12},
  pages={550--574},
  year={Nov. 2024},
  publisher={MDPI}
}

@article{zheng2024convolutional,
  title={Convolutional Neural Network and Ensemble Learning-Based Unmanned Aerial Vehicles Radio Frequency Fingerprinting Identification},
  author={Zheng, Yunfei and Zhang, Xuejun and Wang, Shenghan and Zhang, Weidong},
  journal={Drones},
  volume={8},
  number={8},
  pages={391--411},
  year={Aug. 2024},
  publisher={MDPI}
}

@article{zhou2025cooperative,
  title={Cooperative generative AI for UAV-based scenarios: An intelligent cooperative framework},
  author={Zhou, Longyu and Feng, Wenjiao and Chen, Zihan and Ruan, Tianchen and Leng, Supeng and Yang, Howard H and Fu, Yaru and Quek, Tony QS},
  journal={IEEE Vehicular Technology Magazine},
  year={2025},
  publisher={IEEE}
}
\bibliographystyle{IEEEtran}

\textbf{YAOQI YANG} is an assistant researcher with the National Key Laboratory on Near-Surface Detection, China.

\textbf{YONG CHEN} is a Professor with the Chengdu Fluid Dynamics Innovation Center, China.

\textbf{JIACHENG WANG} is a research fellow with NTU, Singapore.

\textbf{GENG SUN} is a professor with Jilin University, China.

\textbf{DUSIT NIYATO} is a chair professor with NTU, Singapore.

\textbf{ZHU HAN} is a professor with University of Houston, USA.

\end{document}